\documentclass[twocolumn,twoside]{IEEEtran}

\ifCLASSINFOpdf

\else

\fi

\ifCLASSOPTIONcompsoc
    \usepackage[caption=false, font=normalsize, labelfont=sf, textfont=sf]{subfig}
\else
    \usepackage[caption=false, font=normalsize]{subfig}
\fi
\usepackage{lipsum}%
\usepackage[dvipsnames]{xcolor}
\usepackage{algorithm,algorithmic}
\usepackage{balance}
\usepackage{multicol}   
\usepackage{cite}
\usepackage{gensymb}
\usepackage{multirow}
\usepackage{graphics}
\usepackage{epsfig}
\usepackage{graphicx}
\usepackage{epstopdf}
\usepackage{textcomp}
\usepackage{amsmath}
\usepackage{mathtools}
\usepackage{filecontents}
\usepackage{lipsum,color}
\usepackage{amssymb}
\usepackage{float}
\usepackage{colortbl} 
\usepackage{times} 
\usepackage{amsthm}  

\usepackage{amsfonts}

\theoremstyle{break}

\begin{document}

\title{\LARGE VCSEL-Enhanced Holographic Communication for Next-Generation LiFi:\\ State-of-the-Art, Applications, and Future Directions}

\author{Hossein~Safi, {\it Member,~IEEE}, Iman~Tavakkolnia, {\it Senior Member,~IEEE}, and
	Harald~Haas, {\it Fellow,~IEEE}
\thanks{The authors are with LiFi Research and Development Centre, Department of Enigineering, Cambridge University, Cambridge CB3 0FA, UK. (e-mails: {hs905, it360, huh21}@cam.ac.uk). This work was supported by the Future
	Telecoms Research Hubs, Platform for Driving Ultimate Connectivity (TITAN),
	sponsored by the Engineering and Physical Sciences Research Council
	(EPSRC) under Grant EP/X04047X/1 and EP/Y037243/1.} }

\maketitle
\begin{abstract}
Light Fidelity (LiFi) has emerged as a promising wireless technology that exploits the vast unlicensed optical spectrum to complement radio frequency networks. Recent advances in laser-based transmitters, particularly vertical-cavity surface-emitting laser (VCSEL) arrays, enable LiFi systems with multi-gigabit data rates, fine-grained spatial multiplexing, and high energy efficiency. However, the highly directional nature of laser beams introduces new challenges related to user mobility, alignment, and dynamic environments. This article introduces VCSEL-enabled holographic communication as a system-level paradigm that addresses these challenges by tightly integrating communication, sensing, and positioning within a single LiFi architecture. The proposed approach leverages individually addressable VCSEL arrays to form a dense grid of controllable beams, while a real-time digital twin of the environment enables adaptive beam management, environmental mapping through sensing, and user localization through positioning, including non-line-of-sight operation. By tightly integrating high-speed data transmission with environmental perception and user tracking, the LiFi access point evolves from a static transmitter into an intelligent environmental hub. The article also provides a tutorial overview of the underlying hardware, system architecture, and operational principles of holographic LiFi, and discusses key applications, open challenges, and future research directions toward next-generation intelligent optical wireless networks.

\end{abstract}
\IEEEpeerreviewmaketitle

\section{Introduction}
Indoor wireless networks are increasingly expected to support data-intensive and latency-sensitive applications such as immersive extended reality, mobile robotics, and real-time machine control, pushing conventional radio-frequency links to their practical limits \cite{alsaedi2023spectrum}. As future wireless networks evolve toward sixth generation and beyond, it has become increasingly clear that radio technologies alone will struggle to meet the combined demands of capacity, latency, energy efficiency, and reliability. In this context, optical wireless communication, and in particular Light Fidelity (LiFi), has gained renewed attention as a powerful complementary technology that exploits the vast unlicensed visible and infrared spectrum \cite{haas2015lifi}.

Early LiFi systems demonstrated the feasibility of using light emitting diodes for indoor wireless communication, offering advantages such as inherent security, immunity to radio interference, and seamless integration with lighting infrastructure \cite{karunatilaka2015led}. However, the limited modulation bandwidth of LEDs fundamentally constrains achievable data rates, making it difficult for LED based LiFi to support emerging applications such as immersive reality, autonomous systems, and high density industrial connectivity.

To overcome these limitations, the field is rapidly transitioning toward laser based LiFi (also known as LiFi 2.0) to accommodate ultra high data rate applications \cite{soltani2023terabit, 10974735}. In this context, vertical cavity surface emitting lasers (VCSELs) have emerged as a key enabling technology due to their high modulation bandwidth, excellent power efficiency, and compatibility with dense two dimensional array integration \cite{khan2021high}. VCSEL arrays make it possible to generate a large number of narrow, high quality optical beams from a compact transmitter, enabling unprecedented levels of spatial reuse and aggregate throughput. Recent experimental demonstrations have shown that VCSEL-based LiFi systems can achieve tens to hundreds of gigabits per second in indoor environments. Furthermore, it has been experimentally demonstrated that an array of VCSELs can be used to form a grid of beams at the receiver plane, enabling simultaneous multiuser support through a VCSEL-based access point \cite{safi2026chipscale}.

Despite these advantages, laser-based LiFi introduces a fundamental system-level challenge. Unlike LEDs, which provide wide-area illumination, VCSELs emit highly directional beams. While this directionality improves link quality and physical-layer security, it also makes connectivity sensitive to user movement, blockage, and changes in the surrounding environment. In realistic indoor settings with mobile users and dynamic obstacles, maintaining reliable alignment across multiple links becomes challenging, and static or preconfigured beam patterns are insufficient for scalable operation. This motivates a shift toward LiFi systems that are adaptive and environment-aware.

To fully unlock the potential of VCSEL-based LiFi, a shift in system design philosophy is required. Rather than treating the LiFi access point as a passive transmitter, we propose a paradigm in which the network maintains a real-time three-dimensional representation of its environment and actively adapts its optical beams accordingly. We refer to this concept as VCSEL-enabled holographic communication. The term holographic here does not denote classical optical holography, but rather a system-level approach in which communication is continuously informed by an updated digital twin of the physical environment, enabling intelligent beam management and context-aware resource allocation.

In the proposed architecture, an individually addressable VCSEL array forms a dense grid of optical beams that can be dynamically activated, grouped, or redirected. The same optical hardware is also used for environmental sensing, operating in a detection-and-ranging mode to reconstruct the surrounding geometry and identify obstacles, while positioning mechanisms exploit the sensed and communication signals to localize and track users. This enables a unified platform in which beam control and resource allocation are continuously informed by the physical environment. As a result, the LiFi access point evolves into an intelligent environmental hub capable of proactive beam management, non-line-of-sight (NLoS) link establishment, and context-aware resource allocation. While the holographic LiFi paradigm is applicable to a range of optical wireless platforms, this article focuses on VCSEL-based implementations due to their high modulation bandwidth, array scalability, and suitability for fine-grained beam control.

The remainder of this article provides a tutorial overview of this emerging holographic LiFi paradigm. We first introduce the core hardware concept of addressable VCSEL arrays and the grid of beams they enable. We then present the system architecture and closed loop operational principle based on a real-time digital twin. Next, we discuss the benefits of integrated communication, sensing, and positioning, followed by representative application scenarios. Finally, we outline key open challenges and future research directions that must be addressed to realize practical holographic LiFi systems.
\begin{figure}
	\begin{center} 
		\includegraphics[width=3.0 in]{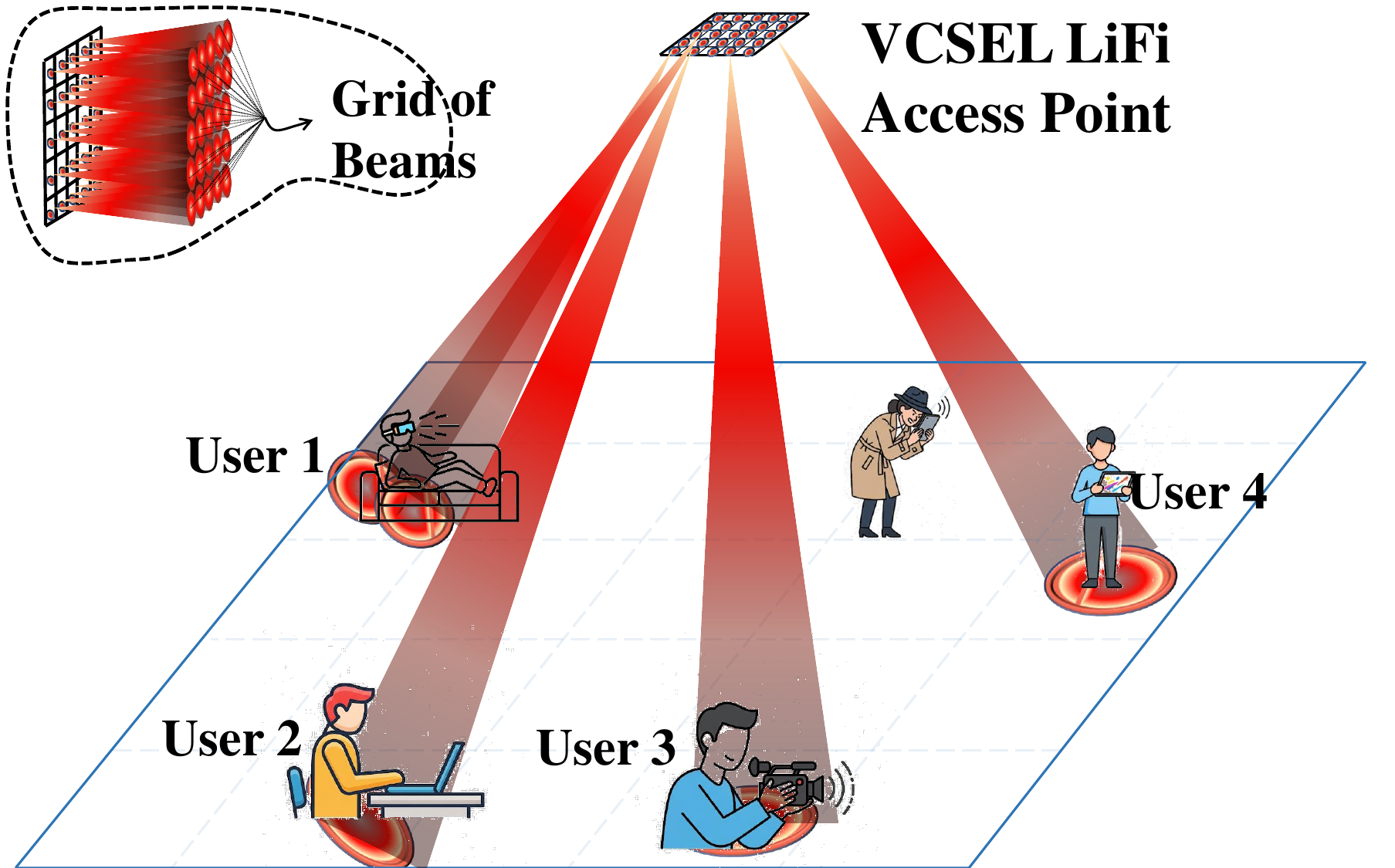}
		\caption{A VCSEL‑array LiFi access point projects a grid of narrow beams indoors, enabling massive spatial multiplexing with dedicated beams for each device. Multiple beams can be combined for high‑rate applications like AR while maintaining security by avoiding illumination of non‑targeted areas.}
		\label{fig1}
	\end{center}
\end{figure}

\section{The Core Hardware: Addressable VCSEL Arrays}

\subsection{Why VCSELs for Next Generation LiFi}
VCSELs emit light perpendicular to the chip surface, which enables compact packaging, on-wafer testing, and high-yield fabrication \cite{ledentsov2022high}. Their far-field radiation patterns are typically near-circular and amenable to precise beam shaping with compact optics, while offering stable wavelength characteristics, making them well suited for both high-speed data transmission and precision optical sensing. From a communication standpoint, VCSELs support multi gigahertz modulation bandwidths, enabling data rates far beyond those achievable with LEDs. Their high power conversion efficiency further supports energy efficient network operation. It was shown that their energy efficiency is approximately twice that of state-of-the-art WiFi links \cite{safi2026chipscale}. These advantages have already established VCSELs as a leading technology in short-reach optical interconnects, are driving their adoption in laser-based LiFi systems, and have led to their widespread use in compact light detection and ranging (LiDAR) modules for sensing and ranging. Equally important is the ability to fabricate VCSELs in dense 2D arrays. Unlike edge‑emitting lasers, VCSELs can be monolithically integrated into compact, individually addressable arrays, a key capability for the concepts discussed here.
\subsection{The Grid of Beams Concept}
An individually addressable VCSEL array transforms the LiFi access point into a programmable grid of narrow optical beams. Each VCSEL element can be activated or deactivated on nanosecond time scales, providing coarse but extremely fast beam steering without mechanical movement \cite{zeng2021vcsel}. This grid of beams also enables fine grained spatial multiplexing. Rather than broadcasting a single beam across an entire room, the access point can establish multiple parallel links, each directed toward a specific user or device (see Fig. \ref{fig1}). Different beams can carry independent data streams, significantly increasing aggregate throughput while minimizing inter user interference. Adjacent VCSEL elements can also be grouped to form composite beams with higher optical power or broader coverage. This allows the system to flexibly adapt beam characteristics to application requirements, for example allocating additional beams to users demanding high data rates, while maintaining efficient use of optical resources \cite{10949139}.

\subsection{Dual Use Operation: Communication and Sensing}
Addressable VCSEL arrays inherently support both data transmission and active environmental sensing using the same optical hardware. In communication mode, selected VCSEL elements are intensity modulated at high speed to transmit data toward a user device with high signal to noise ratio and enhanced physical layer security. In sensing mode, the array operates in a LiDAR like fashion. Short optical pulses are emitted, and the reflected light is captured by a co located photodetector array such as a single-photon avalanche diode (SPAD) array. By measuring time of flight across different beam directions, the system constructs a three dimensional representation of the environment.

These two modes are tightly coupled. Sensing information directly informs beam selection and link management, while communication signals and pilot measurements across multiple beams can be leveraged to refine user localization and tracking. This eliminates the need for separate sensing hardware and ensures that environmental awareness is intrinsically aligned with communication operation. In the next section, we describe how this hardware platform is embedded within a closed-loop system architecture centered on a real-time digital twin of the environment.

\begin{figure*}
	\begin{center} 
		\includegraphics[width=5.1 in]{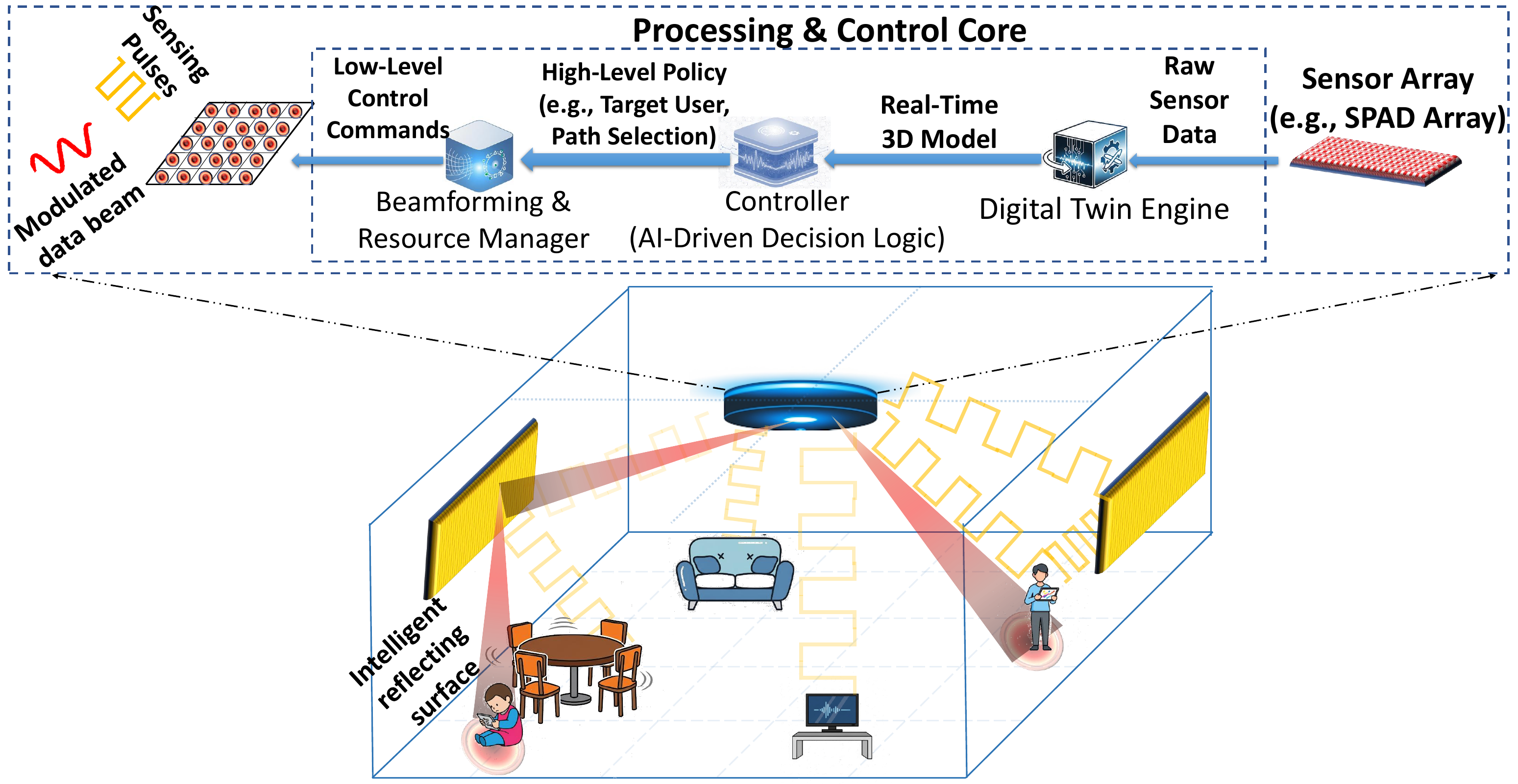}
		\caption{Block diagram of the holographic communication system. The architecture operates in a continuous closed loop. The sensor array captures environmental data, which the Digital Twin Engine uses to build a 3D model. An AI-driven Controller analyzes this model to make strategic decisions, which are then executed by the Beamforming Manager and the VCSEL array. This enables the system to intelligently adapt its communication beams in real-time.}
		\label{fig2}
	\end{center}
\end{figure*}

\section{Holographic System Architecture and Principle of Operation}

With the addressable VCSEL array established as the central hardware component, we now present the system architecture that enables it to operate as an intelligent hub for environmental awareness. The novelty of the holographic communication paradigm lies not just in the hardware, but in the closed-loop, data-driven process that continuously synchronizes the system's actions with the physical world. This is achieved through a cohesive architecture centered on a real-time digital twin as overviewed in the following.

\subsection{System Architecture Overview}
Fig. \ref{fig2} illustrates the proposed architecture. The key functional components of this structure are described as follows.

\subsubsection{VCSEL Array and Driver} As described in Section II, the VCSEL array and its multi channel driver act as the shared optical front end for both communication and sensing. Individual VCSEL elements can be selectively activated and modulated with high temporal precision.

\subsubsection{Sensor Array} Co-located with the VCSEL array is a high-speed, wide-area photodetector array, such as a SPAD array. Its primary role is to capture the reflected light pulses during the sensing phase to enable 3D mapping. For communication, a separate, smaller photodetector at the user equipment receives the modulated data signal from the access point.

\subsubsection{Digital Twin Engine} This is the computational heart of the system. It is a software module responsible for processing the raw sensor data to construct and continuously update a 3D model of the environment. This digital twin contains information about the location of static obstacles (e.g., walls, furniture), dynamic obstacles, and tracked users.

\subsubsection{Controller and Resource Manager} A controller queries the digital twin to obtain real-time environmental context and makes high level decisions related to user scheduling, beam selection, and path planning. These decisions are translated into low level commands by a beamforming and resource manager, which determines which VCSEL elements to activate and how to allocate modulation and coding resources.

\subsubsection{Intelligent Reflecting Surface (Optional Element)}
An intelligent reflecting surface (IRS) can be deployed on selected walls or objects to enhance NLoS connectivity. By providing controllable optical reflections, such surfaces can create additional propagation paths in scenarios where direct LoS links are temporarily blocked. Although not required for the core holographic LiFi concept, IRSs can be integrated into the digital twin to increase path diversity and enhance link robustness in cluttered environments.

\subsection{Closed Loop Operation}
The holographic LiFi system operates in a continuous adaptive loop that allows it to respond rapidly to environmental changes such as user mobility or temporary blockage. This loop consists of four tightly coupled phases.

\subsubsection{Sensing} During the sensing phase, the VCSEL array emits short optical pulses to probe the environment. Reflected photons are captured by the sensor array, enabling coarse three dimensional perception of the surrounding space.

\subsubsection{Modeling} The raw time-of-flight data from the sensor is sent to the digital twin engine. The engine processes this data to update its 3D map, refining the positions of known users and objects and identifying any changes since the last scan.

\subsubsection{Decision} With an updated digital twin, the controller assesses the state of all communication links. For a mobile user, it predicts their trajectory and determines the best VCSEL element to maintain a link. If a direct LoS path is blocked, the controller queries the digital twin for viable reflective surfaces (e.g., mounted on a wall) and calculates the corresponding angles to establish an NLoS link.
\begin{figure*}
	\begin{center} 
		\includegraphics[width=5.3 in]{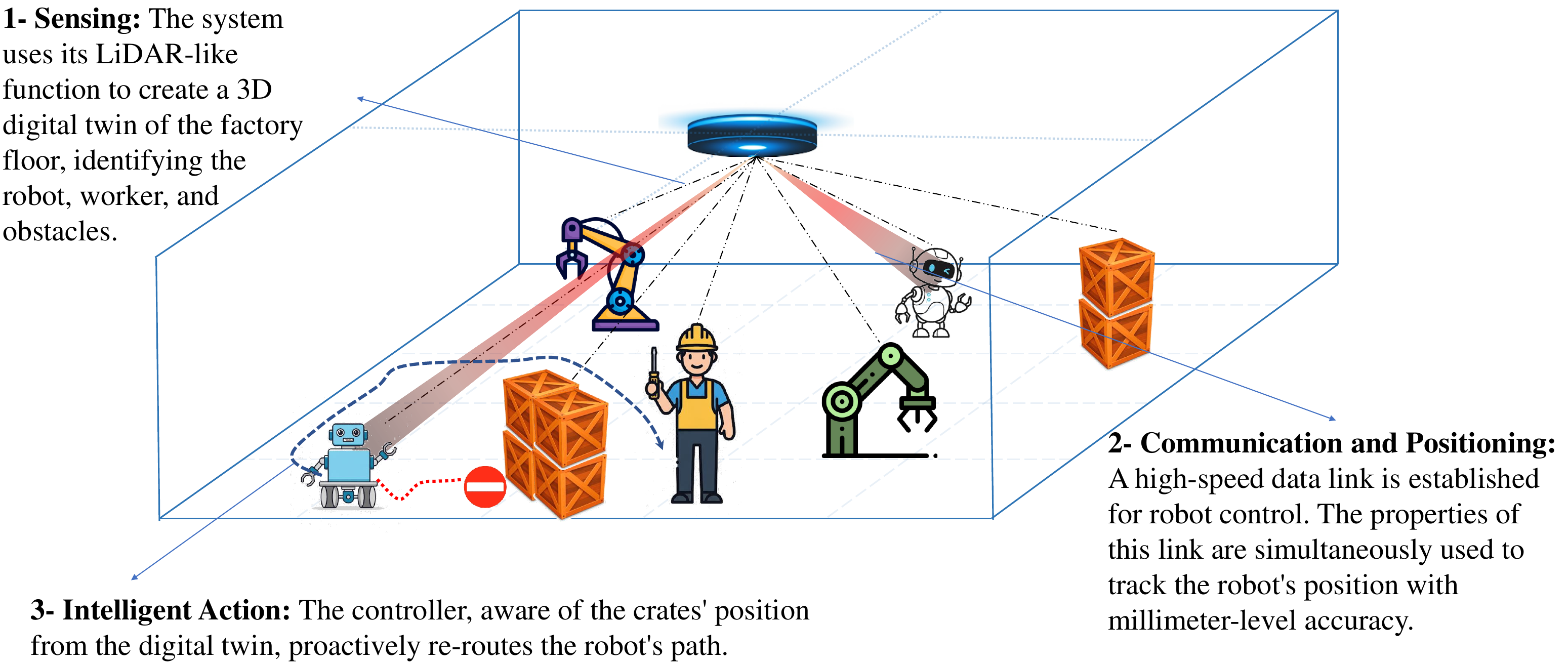}
		\caption{An illustration of the JCSP synergy in a smart factory. The system (1) senses the environment to build a digital twin, (2) establishes a high-speed communication link that also serves for high-precision positioning, and (3) uses this complete environmental awareness to intelligently control the robot.}
		\label{fig3}
	\end{center}
\end{figure*}

\subsubsection{Communication} The controller passes the beam-steering instructions to the beamforming manager, which activates and modulates the chosen VCSEL(s) to transmit data to the user. This communication phase continues until the next sensing cycle is initiated, with the frequency of the loop being dynamically adjustable based on the environment's volatility.

This closed loop operation fundamentally transforms the role of the LiFi access point. Rather than reacting to link degradation after it occurs, the system proactively adapts its beams based on explicit knowledge of the physical environment. Beam management, handover, and NLoS routing are therefore driven by physical awareness rather than purely signal level metrics. As a result, holographic LiFi bridges the gap between communication and environmental perception.

\section{Joint Communication, Sensing, and Positioning}

The holographic communication architecture merges functions that are traditionally handled by separate systems. By using the same VCSEL array for data transmission and environmental sensing, it enables joint communication, sensing, and positioning (JCSP), also known as integrated sensing and communication (ISAC). Here, sensing reconstructs the surrounding geometry and detects obstacles, while positioning leverages sensing outputs and communication measurements to localize and track users. As illustrated in Fig. \ref{fig3}, this creates a closed feedback loop in which improved sensing enhances communication, while communication measurements refine positioning.

\subsection{Sensing Driven Communication Adaptation}
Traditional LiFi systems adapt primarily based on abstract link metrics such as signal to noise ratio or packet error rate, without explicit knowledge of the physical environment. In contrast, holographic LiFi operates with a continuously updated digital twin, allowing the network to associate performance changes with their physical causes. For example, if a link degrades, the controller can determine whether the cause is user movement, temporary blockage, or changes in the reflective environment. Rather than increasing transmit power, the system can proactively switch beams, reallocate optical resources, or establish a NLoS path via a known reflective surface. 

\begin{figure}[!t]
	\centering
	\includegraphics[width=\columnwidth]{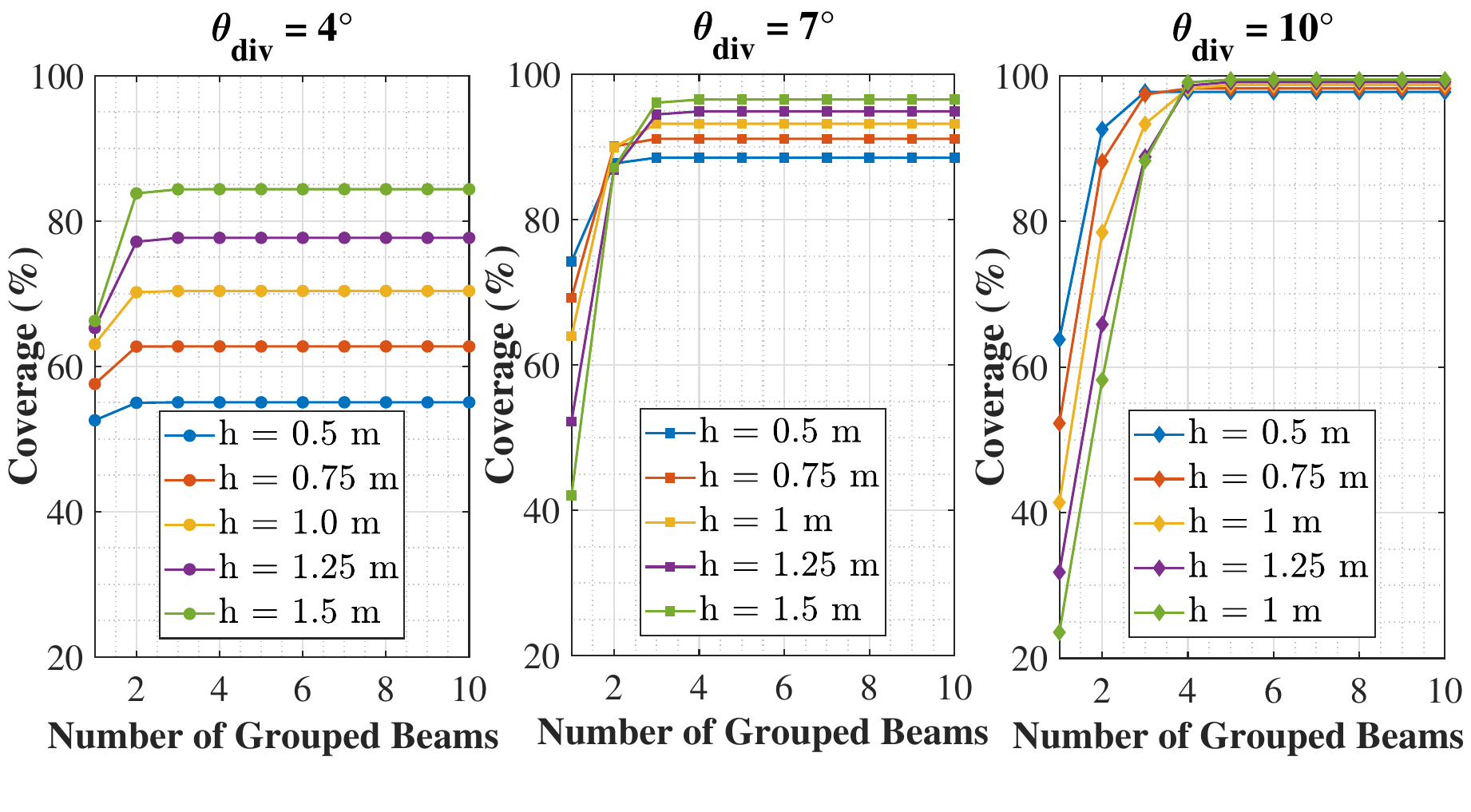}
	\caption{Coverage performance versus the number of grouped beams for different divergence angles and receiver heights based on the link parameters given in \cite{11152841}. Beam grouping improves coverage for moderate divergences, though large group sizes show diminishing returns. This demonstrates how adaptive grouping and divergence control provide robust 3D coverage with minimal optical resources.}
	\label{fig:3d_coverage}
\end{figure}

This context-aware adaptation improves link robustness and enables more predictable quality of service, while reducing unnecessary optical transmit power at the cost of additional sensing and computation overhead, which must be balanced against the energy savings achieved at the transmitter. To provide quantitative insight into the benefits of adaptive beam control, Fig. \ref{fig:3d_coverage} presents a simple coverage analysis based on a grid of VCSEL beams in an indoor environment. The figure shows the coverage percentage as a function of the number of grouped beams for different beam divergence angles and receiver heights. For narrow beams, coverage remains limited even when multiple beams are grouped, indicating that insufficient spatial spread cannot be compensated by beam aggregation alone. In contrast, moderate divergence combined with beam grouping rapidly achieves high coverage across the three-dimensional user space, while larger divergence angles reach near-full coverage with only a small number of grouped beams. Beyond this point, further beam grouping yields diminishing returns, highlighting that environment-aware controllers can balance divergence and beam grouping to maintain robust coverage without activating the full VCSEL array.

\vspace{-3mm}

\subsection{Communication-Assisted Positioning and Unified Environmental Interaction}

Holographic LiFi dissolves the traditional separation between communication, sensing, and positioning by using a shared optical front end for all three functions. While LiDAR-like sensing provides coarse environmental mapping and obstacle detection, measurements extracted from communication signals across multiple beams, such as relative received power, angular diversity, and timing information from synchronization sequences, can be exploited to refine user localization and tracking with high spatial precision  \cite{kouhini2021lifi}. This joint operation transforms the optical access point into a unified interface between the network and its physical environment, enabling the simultaneous delivery of multi-gigabit data rates, user tracking, and environmental awareness within a single platform.

This unified approach reduces hardware redundancy and simplifies system design, while also aligning with broader trends in integrated sensing and communication. It positions holographic LiFi as the natural optical counterpart to radio-based ISAC frameworks. These capabilities are particularly valuable for applications that require tight coordination between connectivity and spatial awareness, such as immersive extended-reality systems and industrial automation. The following section examines how these capabilities translate into practical application scenarios.

\begin{figure*}
	\begin{center} 
		\includegraphics[width=5.4 in]{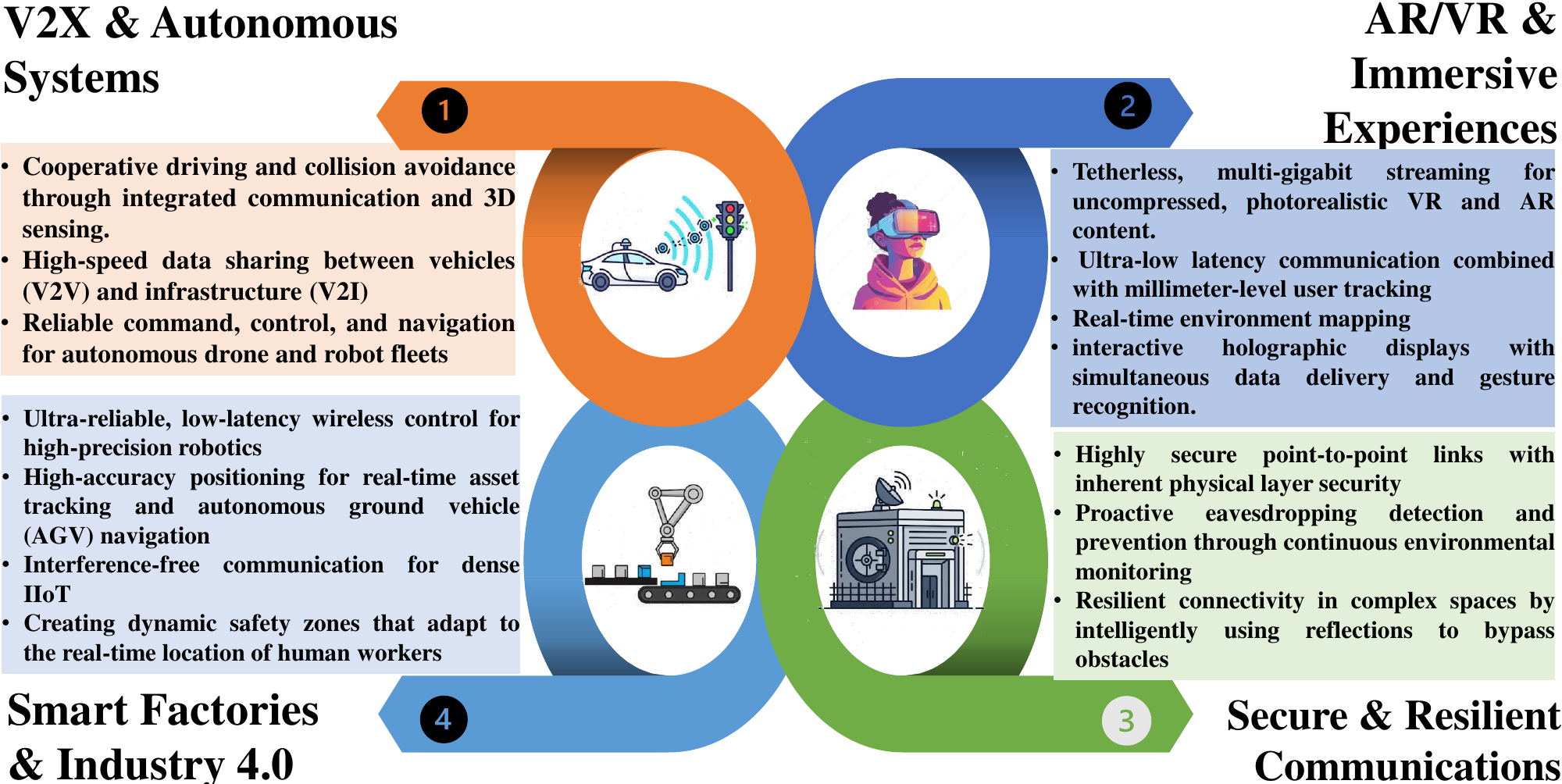}
		\caption{A summary of key application domains where VCSEL-enhanced holographic LiFi can have a transformative impact. These include (1) autonomous systems and V2X networks, (2) immersive AR/VR experiences, (3) smart factories and industrial automation, and (4) high-security communication environments.}
		\label{fig4}
	\end{center}
\end{figure*}
\section{Applications and Use Cases}
By combining high speed optical communication with real-time sensing and positioning, VCSEL enabled holographic LiFi enables applications that are difficult to support with conventional wireless technologies (see Fig. \ref{fig4}). The ability to deliver multi gigabit data rates while maintaining fine grained spatial awareness makes this paradigm particularly well suited for dynamic, safety critical, and data intensive environments.

\subsection{Autonomous Systems and Vehicular Networks}

\subsubsection{Vehicle-to-Everything (V2X) Communication} In vehicular networks, the holographic LiFi system can provide ultra-high-bandwidth links between vehicles (V2V), infrastructure (V2I), and pedestrians (V2P). This enables the sharing of high-resolution sensor data, cooperative maneuvers, and real-time traffic updates. Simultaneously, the integrated LiDAR functionality provides a detailed 3D map of the immediate surroundings, acting as a primary or secondary sensor for collision avoidance and navigation, even in GPS-denied environments like tunnels or urban canyons.

\subsubsection{Drone and Robotics Coordination} In indoor and semi indoor settings such as warehouses, factories, and logistics hubs, the same infrastructure can coordinate fleets of robots or drones. Dedicated optical beams support low latency command and control, while the digital twin enables safe navigation, collision avoidance, and cooperative task execution without reliance on external positioning systems.

\subsection{Immersive Augmented and Virtual Reality (AR/VR)}

\subsubsection{Tetherless High-Fidelity VR} Truly immersive AR/VR experiences demand both enormous bandwidth and ultra-low latency to prevent motion sickness and create a sense of presence. Also, current high-end VR headsets are often tethered by cables to a powerful computer. Holographic LiFi can cut this cord, wirelessly streaming uncompressed video data to the headset while the integrated positioning system tracks the user's head and body movements with millimeter-level accuracy. This combination of high throughput and precise tracking is the key to a truly mobile and believable VR experience.

\subsubsection{Environment-Aware AR} For augmented reality, the system cannot only deliver data to AR glasses but also use its sensing capabilities to map the user's environment in real-time. This allows digital information to be correctly anchored to real-world objects, thereby creating a meaningful and interactive experience.

\subsection{Smart Environments: Factories and Healthcare}

\subsubsection{Industry 4.0} In advanced manufacturing facilities, the system can provide the robust connectivity needed for the industrial internet of things (IIoT). It can control machinery with high precision while simultaneously monitoring the factory floor for safety, tracking assets, and guiding autonomous ground vehicles (AGVs).

\subsubsection{Telemedicine and Smart Hospitals} The high-resolution imaging and zero-latency communication offered by this technology can revolutionize remote healthcare. It can enable surgeons to perform remote operations using robotic arms, with real-time 3D holographic feedback of the patient. In a hospital, the system can provide secure, high-speed data access while tracking medical equipment and monitoring patient vital signs without contact.

\subsection{Secure and Resilient Communication}
The highly directional nature of the VCSEL beams provides a significant enhancement to physical layer security. It is extremely difficult to eavesdrop on a narrow beam of light without physically obstructing it, an act that would be immediately detected by the system's sensing capabilities. This makes the technology ideal for applications requiring high security, such as financial institutions, military operations, and corporate headquarters.

\section{Open Challenges and Future Research Directions}

While VCSEL enabled holographic LiFi offers a compelling vision for next generation optical wireless networks, several challenges must be addressed before large scale deployment becomes practical. These challenges span hardware, computation, and system intelligence, and they define key directions for future research endeavours.

\subsection{Hardware and Component Evolution}
The performance of the entire system is fundamentally tied to the capabilities of its core components, primarily the VCSELs themselves. Future research must focus on:

\subsubsection{VCSEL Performance and Noise Reduction} Pushing data rates toward Tb/s will require further reductions in intrinsic noise and improved thermal stability of VCSELs. Optimizing the laser's design, improving material purity to reduce defects, and developing sophisticated electronic feedback systems to stabilize the driving current are critical research paths to enhance the signal quality for both communication and high-precision sensing \cite{ledentsov2022high}.

\subsubsection{Advanced Laser Architectures} The coherent nature of laser light can create speckle, a granular noise pattern that can degrade sensing and imaging performance. Exploring novel laser designs, such as chaotic-cavity surface-emitting lasers (CCSELs), which are engineered to have reduced spatial coherence, could mitigate speckle noise. Although currently less mature than VCSELs, CCSELs remain a promising research direction for applications that demand high‑fidelity imaging \cite{alkhazragi2023modifying}.

\subsubsection{Sensor Technology} The development of larger, faster, and more sensitive photodetector arrays (e.g., SPADs) is crucial for rapid, accurate construction of the digital twin, especially in low-light conditions or at longer ranges.

\vspace{-4mm}

\subsection{Real-Time Digital Twin Construction}
The concept of a continuously updated digital twin is central to the proposed architecture, but it also presents significant computational hurdles. Future researches should cover the following challenges:

\subsubsection{Computational Complexity} Processing the vast amount of data from a sensor array to generate and update a high-fidelity 3D model in real-time is computationally intensive. Future work must focus on developing highly parallelized algorithms and potentially leveraging dedicated hardware accelerators (e.g., FPGAs) to meet the stringent latency requirements of mobile applications.

\subsubsection{Sensor Fusion and Synchronization}
Robust digital twins require fusing data from multiple sources, such as optical sensors, cameras, and intelligent reflecting surfaces, into a coherent world model. This calls for sensor fusion algorithms that can adaptively weight heterogeneous measurements and remain robust to noise sources common in optical systems, including background noise, shot noise, and speckle. Cross-validation across modalities can mitigate unreliable measurements, while accurate time synchronization across sensing and communication links is essential to maintain consistency in the reconstructed environment.
\begin{figure*}
	\begin{center} 
		\includegraphics[width=5.4 in]{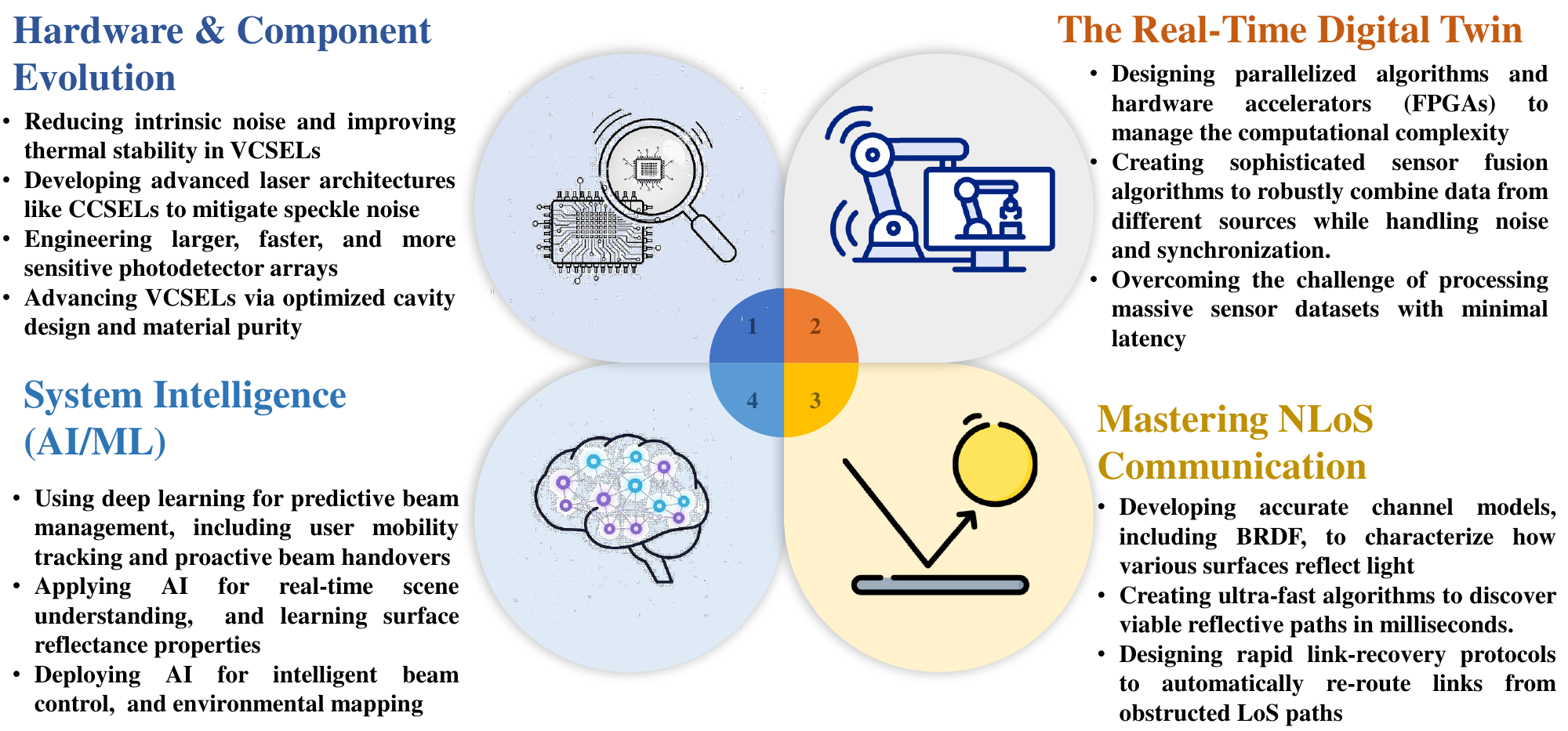}
		\caption{A summary of key research challenges and future directions. Progress in these four areas, i.e., (1) hardware evolution, (2) real-time digital twin computation, (3) NLoS link reliability, and (4) AI-driven system intelligence, is critical for advancing holographic LiFi systems.}
		\label{fig5}
	\end{center}
\end{figure*}

\subsection{AI Driven System Intelligence}
Artificial intelligence and machine learning (AI/ML) are essential tools for managing the complexity of a dynamic holographic LiFi system. Future research will be pivotal in:

\subsubsection{AI-Driven Beam Management} Using AI to manage the grid of beams is a primary research avenue. Deep learning models can be trained to predict user mobility and proactively hand over beams, calculate optimal NLoS paths in real-time, and allocate network resources to maximize throughput and minimize interference in dense user environments.

\subsubsection{AI-Based Environment Mapping and Sensing}

The holographic LiFi system benefits from integrating AI-enabled scene understanding to adapt in real-time. Computer vision and AI can create real-time 3D maps of the indoor space using camera/LiDAR data. Also, AI models can learn how various surfaces reflect light, enabling intelligent exploitation of NLoS paths for coverage extension. Furthermore, AI can detect and track transient obstructions (e.g., people walking) and trigger rapid beam adjustments.

\subsubsection{AI-Enhanced Positioning}
AI-based models can improve positioning accuracy and robustness in VCSEL-based holographic LiFi by learning the relationship between beam-level measurements and user location within the digital twin. By fusing features such as relative received power across the VCSEL beam grid, angular diversity, and timing information from communication pilots, learning-based approaches can compensate for multipath reflections, partial blockage, and device orientation effects. This enables more reliable user tracking in dynamic indoor environments, particularly when geometric sensing alone provides limited resolution.
\subsection{Reliable NLoS Operation}
Although exploiting reflections is a key advantage, ensuring reliable non-line-of-sight links remains challenging due to strong dependence on surface material, texture, and geometry. In addition to passive reflections, intelligent reflecting surfaces can be integrated into the digital twin to provide more controllable NLoS paths in challenging environments. Key research directions include the following.

\subsubsection{Channel Characterization and Modeling}
Accurate channel models are needed for reflective surfaces such as walls, ceilings, windows, and engineered reflectors, including bidirectional reflectance distribution function models tailored to LiFi that capture angle of incidence, wavelength, and surface roughness. Models for intelligent reflecting surfaces should also capture the impact of surface configuration on reflected fields.

\subsubsection{Reliable NLoS Link Establishment}
Algorithms are needed to rapidly identify viable reflective paths using the digital twin and establish reliable links under dynamic conditions. This includes exploiting both passive reflections and configurable intelligent reflecting surfaces, combining signals from multiple reflection points for robustness, and enabling fast path discovery and link recovery through ray tracing or AI-accelerated digital twin querying.
\vspace{-2mm}

\section{Conclusion}

This article has presented VCSEL-enabled holographic communication as a system-level paradigm for intelligent, environment-aware LiFi networks. By combining addressable VCSEL arrays with a real-time digital twin, the LiFi access point evolves into a context-aware optical node that jointly supports communication, sensing, and positioning. This enables proactive beam control, precise localization, and resilient connectivity in dynamic environments, with applications ranging from autonomous systems to immersive extended reality and industrial networking. While challenges remain in hardware integration, real-time processing, and AI-assisted control, ongoing advances point to a clear path forward. Holographic LiFi thus represents a promising foundation for next-generation intelligent optical wireless networks.

\bibliographystyle{IEEEtraN}
\balance
\bibliography{IEEEabrv,myref}

\end{document}